\documentclass[11pt,a4paper]{article}
\pdfoutput=1

\usepackage{jheppub}

\usepackage{booktabs}
\usepackage{array}
\usepackage{amsmath}
\usepackage{amssymb}
\usepackage{graphicx}
\usepackage{float}
\usepackage{braket}
\usepackage{hyperref}
\usepackage{slashed}
\usepackage{enumerate}
\usepackage[all]{hypcap}
\usepackage{feynmp}

\def\beq{\begin{gather}}
\def\eeq{\end{gather}}
\def\beeq{\begin{eqnarray}}
\def\eeeq{\end{eqnarray}}

\def\vec#1{\mbox{\boldmath $#1$}}

\def\gp2{g^{\prime 2}}

\usepackage{color}

 \newcommand{\column}[1]{\left(\begin{array}{c} #1 \end{array}\right) }

\title{On the Structure of Anomalous Composite Higgs Models} 
\date{\today}

\preprint{Cavendish-HEP-16/09, DAMTP-2016-41}

\author[a]{Ben Gripaios,}
\author[a,b,c]{Marco Nardecchia,} 
\author[a,b]{Tevong You}

\affiliation[a]{\normalfont{Cavendish Laboratory, University of Cambridge, 
J.J. Thomson Avenue, Cambridge, CB3 0HE, UK}}
\affiliation[b]{\normalfont{DAMTP, University of Cambridge, 
Wilberforce Road, Cambridge, CB3 0WA, UK}}
\affiliation[c]{\normalfont{CERN, Theoretical Physics Department, Geneva, Switzerland}}

\emailAdd{gripaios@hep.phy.cam.ac.uk} 
\emailAdd{m.nardecchia@damtp.cam.ac.uk}
\emailAdd{tty20@cam.ac.uk}

\abstract{
We describe the anomaly structure of an composite Higgs model in which the $SO(5)/SO(4)$ coset
structure of the minimal model is extended by an additional,
non-linearly-realized $U(1)_{\eta}$. In addition, we show that the effective
  lagrangian admits a term that, like the
  Wess-Zumino-Witten term in the chiral lagrangian for QCD, is not
  invariant under the non-linearly realized symmetries, but rather
  changes by a total derivative. This term is unlike the
  Wess-Zumino-Witten term in that it does not arise from anomalies.
  If present, it
may give rise to the rare decay $\eta \rightarrow h W^+ W^- Z$. The
phenomenology of the singlet in this model differs from that in a
model based on
$SO(6)/SO(5)$, in that couplings to both gluons and photons, arising
via anomalies, are present. We show that while some tuning is needed to accommodate flavour and electroweak precision
constraints, the model is no worse than the minimal model in this regard. 

}  

\begin{document}   
 
\maketitle 

\section{Introduction}
\label{sec:intro}

In recent years, theorists have devoted much attention to models in
which the electroweak hierarchy problem is solved by postulating
that the Higgs boson arises as a composite pseudo-Goldstone boson of
some new,  TeV-scale strong dynamics
\cite{Kaplan:1983fs,Georgi:1984af,Dugan:1984hq}. 

If this is really what happens in Nature, then it is interesting to
ask how we might go about figuring out what the underlying UV dynamics
is, given our current rather poor theoretical understanding of
strongly-coupled dynamics. 

One way in which may we may do so is via triangle anomalies, which are
not renormalized and so, if present in the UV, must be reproduced in
the IR, either by massless fermions or by terms involving the pseudo-Goldstone bosons. Such anomalies are
not only not renormalized, but they are also topological in nature. This means that by
measuring them in the IR, we may gain concrete information about
the UV dynamics. The classic example, of course, is in QCD, where the
measurement of the decay rate $\pi^0
\rightarrow \gamma \gamma$ (which arises via the electromagnetic
anomaly \cite{Bell:1969ts,Adler}) enables us to
infer that $N_c =3$. 

In order to make such spectacular inferences, one must be lucky enough
to have a low-energy lagrangian that admits a non-trivial anomaly
structure. The minimal, and by far the most popular, composite Higgs
model, based on $SO(5)/SO(4)$ \cite{Agashe:2004rs} does not feature
anomalies. However, the `next-to-minimal' model based on $SO(6)/SO(5)$
\cite{Gripaios:2009pe}, which is just as good from the
phenomenological point of view, does. Compared to the minimal model,
it features only an additional electroweak singlet scalar, which couples to
electroweak gauge bosons via a single $SO(6)^3$ triangle anomaly. 

Here we wish to describe yet another model, based on
$SO(5)\times U(1)/SO(4)$. It is just as minimal as the $SO(6)/SO(5)$ model,
in the sense that it features only an additional electroweak singlet
scalar. But it turns out to have a much richer anomaly
structure, with several novel features.

A first novel feature is that there are now 3 distinct triangle anomalies, which
give rise, at leading order, to couplings of the singlet to both gluons and electroweak
bosons. 

A second novel feature is that the higher-order structure of the
anomalous effective action is not unique. Indeed, we exhibit two
solutions to the Wess-Zumino consistency conditions. As far as we are
aware, this phenomenon has not been observed before in the literature
on sigma models.

A third novel feature is that the 
effective lagrangian admits a term that
  is not invariant, but rather changes by a total derivative, under the
  non-linearly realized symmetries. Such a term is much like the
  Wess-Zumino-Witten (WZW) term in the chiral lagrangian of QCD, which allows processes violating a putative internal symmetry under which
Goldstone bosons change sign, such as $K+\overline{K} \rightarrow 3\pi$
\cite{Wess:1971yu,witten}. But there is one noteworthy distinction
between the WZW-like term
presented here and WZW term in QCD. In the latter, the presence of the anomaly implies
the presence of the WZW term, in the sense that the
low-energy effective action reproducing the anomaly
reduces to the WZW term when the gauge fields vanish. In the model
presented here, this is
not so. This phenomenon is also, we believe,
unknown in the sigma-model literature. 
The WZW-like term is also of phenomenological interest, in that it
may lead to a rare decay of the singlet via
  $\eta \rightarrow h W^+ W^- Z$.

The outline is as follows.  In the next Section, we present the
pattern of symmetry breaking and sketch the concomitant
anomalies. We then present a
full discussion of the anomaly structure and the WZW-like term in \S \ref{sec:wzw}.
In \S \ref{sec:ferm}, we describe the couplings to fermions and
the implications for flavour physics. In \S \ref{sec:pot}, we
discuss the form of the scalar potential that is induced by the
couplings to gauge fields and fermions. We conclude in \S \ref{sec:conclusion}. Two more technical discussions are relegated to appendices.

\section{The model}
\label{sec:ch}
We wish to consider composite Higgs models based on a homogeneous space $G/H$
that feature triangle anomalies.\footnote{See \cite{Gripaios:2008ei,Gripaios:2009pe} for
  earlier discussions of anomalies in composite Higgs models.}

The minimal model \cite{Agashe:2004rs}, based on $SO(5)/SO(4)$ (or
$SO(5)/O(4)$ with custodial protection of $Z\rightarrow b\overline{b}$ \cite{Agashe:2006at}), features
no triangle anomalies. The `next-to-minimal' model based on
$SO(6)/SO(5)$ \cite{Gripaios:2009pe} does, however, feature triangle
anomalies. Indeed the Goldstone bosons transform as the 5-d irrep of
$SO(5)$, which, on restriction to the 
$SO(4)$ subgroup, yields both a 4-d irrep (viz. the
Higgs field) and a singlet. Moreover, since $SO(6)$ is locally
isomorphic to $SU(4)$, we have the possibility of an $SU(4)^3$
triangle anomaly.\footnote{Since $H_{dR}^5 (SO(6)/SO(5)) = H_{dR}^5 (S^5)
= \mathbb{R}$, there is also a possible WZW term.  As explained in the next sections, $H_{dR}^5$ denotes the fifth de Rham cohomology group. } 
This anomaly leads to an interaction, at leading order, of the form $\frac{1}{16 \pi^2} \frac{\eta}{f} (g^2_2 \, W_{\mu \nu}\tilde{W}^{\mu \nu}- g_1^2 \, B_{\mu \nu}\tilde{B}^{\mu \nu})
$, with neither a coupling to gluons nor to photons \cite{Bellazzini:2015nxw}. 

The absence of a coupling to photons in this model is something of a
group-theoretical accident, in that there are couplings to $ZZ$, $\gamma Z$,
and $WW$. But the absence of a coupling to gluons looks, at first
sight, to be a generic problem in a composite Higgs model, given that the r\^{o}le of the new strong dynamics is
to break the electroweak symmetry, 
independently of the $SU(3)_C$ dynamics. In fact, this is not so, since a consequence of partial
compositeness is that the new strong sector must be charged under $SU(3)_C$
\cite{Gripaios:2009dq}. So it seems quite plausible that the elementary fermions
of the UV theory could generate an anomaly involving $SU(3)_C$. 

One way to get couplings of the singlet to both electroweak
gauge bosons and to gluons via anomalies is to include both $SU(3)_C$ and
$SU(2)_L$ or $U(1)_Y$ in some simple subgroup of $G$. But such a strategy
will lead to additional coloured Goldstone bosons, with
potentially dangerous phenomenological implications.\footnote{Such
  states may also have desirable phenomenological implications,
  however \cite{Gripaios:2010hv,Gripaios:2014tna}.}
A safer, and simpler, strategy is to modify the minimal model by adding a
non-linearly realized
$U(1)_{\eta}$ factor,
such that the symmetry breaking pattern in the strong sector becomes 
\begin{equation} \label{eq:gh}
\frac{G}{H} = \frac{SU(3)_C \times SO(5) \times U(1)_{X} \times U(1)_{\eta} }{SU(3)_C \times SO(4) \times U(1)_{X}} \, ,
\end{equation}
where $U(1)_X$ denotes the usual $U(1)$ needed in composite Higgs models to give the
correct hypercharge assignments to SM fermions. This model features an additional SM singlet compared to the minimal
composite Higgs model. We remark that, unlike the $SO(6)/SO(5)$ model,
this coset space allows for two distinct decay
constants, $f$ and $f_{\eta}$, associated with the Higgs boson and the
$\eta$, respectively. We assume henceforth that these are generated by the
same strong dynamics, and hence are of the same order of magnitude.

Let us now consider the possible triangle anomalies in this model. As
we shall see in \S \ref{sec:wzw},
triangle anomalies in $G$ are admissible
only if they vanish on restriction to $H$. Thus, our model admits
3 possible sources of triangle anomalies, namely
$SU(3)_C^2 \, U(1)_{\eta}$ and $SO(5)^2 \, U(1)_{\eta}$ anomalies, and anomalies involving $U(1)_{\eta}$ and $U(1)_X$.

The leading contributions to the resulting low-energy effective action arise at
dimension-5, taking the form
\begin{equation}
\label{eq:5d}
\mathcal{L}_{eff} = \frac{1}{16 \pi^2} \frac{\eta}{f_\eta} (c_3 \, g^2_3 \,  G_{\mu \nu} \tilde{G}^{\mu \nu} + c_5 ( g^2_2  W_{\mu \nu} \tilde{W}^{\mu \nu}+ g^2_1  B_{\mu \nu} \tilde{B}^{\mu \nu} ) +c_1 g^2_1  B_{\mu \nu} \tilde{B}^{\mu \nu}),
\end{equation}
where the coefficients are real, but otherwise arbitrary (corresponding to the freedom to
arbitrarily choose the $U(1)_\eta$ irreps of fermions in the UV
theory that contribute to the anomaly). 

\section{Anomalies and WZW-like terms}
\label{sec:wzw}
We now discuss the anomaly structure of the model in more detail,
together with the phenomenological consequences. Let us begin with a
general discussion.
A theory with internal global symmetry group $G$ may be anomalous, in
the sense\footnote{We consider only triangle anomalies here.} that there is no way to regularise the theory such that
the divergences of 3-point functions of conserved currents are all
vanishing. Such anomalies are not renormalized and must be reproduced
at all energies, with consequences for low-energy physics. 

One consequence is a consistency condition on the possible pattern of
symmetry breaking at low energy: if a subgroup $H \subset G$ is
linearly realized at low energy, then $H$ must be anomaly free.
The reason \cite{Preskill:1990fr} is that we could perturb the theory
in an
arbitrarily 
small way by gauging the whole of $G$, but choosing the gauge coupling to
be arbitrarily small. If there were anomalies in $H$, the
gauge bosons in $H$ could get masses via a loop diagram formed out of
two anomalous vertices, implying that $H$ could not 
be linearly realised. 

Once this restriction has been taken into
account, it can be shown that the remaining anomalies can be
reproduced satisfactorily at low-energies
by Goldstone boson
contributions \cite{Chu:1996fr} and an
 explicit formula for the anomalous contribution to
the low-energy effective action for a reductive
homogeneous space $G/H$
can be found (see also \cite{Weinberg}).
As in \cite{Chu:1996fr}, in this Section we employ the langauge of differential forms
and omit normalization factors, giving the result only for the special
case of a symmetric space, which is sufficient for our needs. The formula is most conveniently
written in the fully-gauged case; the result for gauging a subset $F \subset
G$ can be obtained by setting the corresponding gauge fields to zero in the
formula.\footnote{We caution the reader that the symmetry group of the resulting theory is not $G$, even at the
classical level, but rather is the
  normalizer of $F$ in $G$ \cite{Gripaios:2015qya}.} Let $\mathfrak{g}$ and $\mathfrak{h}$ be the Lie
algebras of $G$ and $H$. Since $G/H$ is reductive and symmetric, $\exists$ a split, $\mathfrak{g} = \mathfrak{h} + \mathfrak{k}$, such that
$[\mathfrak{h},\mathfrak{k}] \subseteq \mathfrak{k}$ and
$[\mathfrak{k},\mathfrak{k}] \subseteq \mathfrak{h}$, together with an `internal parity' automorphism
of $\mathfrak{g}$ given by $\mathfrak{h} \rightarrow \mathfrak{h}$ and
$\mathfrak{k} \rightarrow -\mathfrak{k}$.
Letting $A$ be a
$\mathfrak{g}$-valued 1-form representing the gauge fields and letting the coset representative be $e^\xi$,
with $\xi \in \mathfrak{k}$, we have that
\begin{gather} \label{eq:Gamma}
W[\xi,A] = \sum_{\pm} \int_0^1 dt \int d^4x c_\pm \mathrm{tr} [\xi G^\pm [A_t]],
\end{gather}
where $A_t = e^{t\xi} (A+d) e^{-t\xi} \implies F_t = e^{t\xi} F
e^{-t\xi}$, $c_\pm$ are arbitrary coefficients and
\begin{align}
G^+[A] &= 3F_h^2 + F_k^2 -4(A_k^2 F_h +A_k F_h A_k + F_h A_k^2) +
         8A_k^4,\\
G^-[A] &=\frac{3}{2}(F_hF_k + F_k F_h - F_kA_k^2 -A_k F_k A_k -A_k^2 F_k).
\end{align}
Here, $G^\pm$ are the
positive/negative eigenstates with respect to the internal parity and the subscripts $h$ and $k$ denote projections onto the
corresponding subspaces, such that
$F_h = dA_h + A_h^2 +A_k^2, 
F_k = dA_k + A_hA_k +A_k A_h$.

The action (\ref{eq:Gamma}) is unique in the sense that it is the only
action which
vanishes when the Goldstone bosons vanish and whose anomaly is given by $\delta_\alpha \Gamma = \sum_\pm{c_\pm
  \mathrm{tr} \alpha G^\pm [A]}$
\cite{Weinberg}. But it is not unique in the sense that the anomaly
can take many forms, corresponding to the addition of local
counterterms to the effective action. (For a counterexample, it
suffices to choose $H = 0$, for which {\em any} form $G[A]$ for the
anomaly is reproduced by the effective action $\Gamma = \int_0^1 dt \int_x \mathrm{tr} \xi G [A_t]$.)
The action (\ref{eq:Gamma}) is
the one obtained by starting from the canonical form of the anomaly
(which is symmetric with respect to $G$) and subtracting a counterterm
that enforces the vanishing of the anomaly on $H$
\cite{Chu:1996fr}. Hadronic data suggest that this is the option
chosen by the strong interactions, but we are unaware of an argument
that it is the only consistent option.

Even though its {\em raison d'\^{e}tre} is to reproduce anomalies that
arise due to gauging, (\ref{eq:Gamma}) may not vanish in the limit
that gauge fields vanish. In that limit, we obtain 
\begin{gather}
W[\xi,0] = \int_0^1 dt \int d^4x c_+ \mathrm{tr} [\xi (e^{t\xi}de^{-t\xi})_k^4].
\end{gather}
Such a term, which contains an undifferentiated
Goldstone boson at leading order
is not invariant under a $G$ transformation, but rather changes by a total
derivative. We will call such non-invariant lagrangian terms `WZW
terms', in honour of their prototype in the chiral lagrangian.  It was shown in
\cite{D'Hoker:1994ti} that for compact $G$ in $d=4$, and for field
configurations in the trivial fourth homotopy class,\footnote{As usual, we identify spacetime, with fields thereon tending to
a constant value at infinity, with $S^4$.} such terms are in 1-1 correspondence with the
generators of the fifth de Rham cohomology group of
$G/H$. 

We caution the reader that not all such terms can arise from
effective actions reproducing triangle anomalies. By way of a counterexample,
consider the homogeneous space
$SU(2)\times SU(2)/U(1)$, where the $U(1)$ is included in one of the
$SU(2)$s. This space is equivalent as a smooth manifold to $S^3 \times S^2$ and a
straightforward generalization of the arguments presented below shows
that $H^5_{dR} (S^3 \times S^2)  = \mathbb{R}$. Thus, there is a WZW
term in this case, but since $SU(2)$ has no triangle anomalies, it
cannot arise from reproducing them.\footnote{Moreover, since $\pi_4 (S^3 \times S^2) = \mathbb{Z}/2$, one
  cannot use Witten's trick to write the WZW term as an integral over a
  5-disk in this case.} 
\subsection*{Composite Higgs model anomalies} 
For the $SO(5)\times SU(3) \times U(1) \times U(1) / SO(4) \times
  SU(3) \times U(1)$ model,
it is straightforward to
  show that the effective action (\ref{eq:Gamma}) reduces, at leading
  order, to (\ref{eq:5d}).
For $SU(3)^2
U(1)$ and the anomalies involving $U(1)$s, there are no higher-order
corrections to the effective action. There are, however, higher-order corrections for the
$SO(5)^2 U(1)$ anomaly, the detailed calculation of which we relegate
to Appendix \ref{app:anlo}. The next-to-leading order corrections arise at
dimension 7, up to which order the effective action is given, in the
operator basis of \cite{Gripaios:2016xuo}, by
\begin{gather} \label{eq:effactho}
\int c_5 \eta (W^iW^i + B^2 -\frac{16}{9f^2}(H^\dagger H (W^i W^i + B^2) + 2 H^\dagger
\sigma^i H W^i B))+\dots,
\end{gather}
where $W^i$ and $B$ are the field strength 2-forms and $f$ is the
non-linear scale.

These corrections to the leading-order action appear to constitute a
definite prediction of the model, once $c_5$ has been determined from
measurements at leading-order. Unfortunately, the issue of
non-uniqueness discussed above now rears its ugly head. Indeed, it is
easy to check that the 
the leading-order action (\ref{eq:5d}) alone also provides a
solution of the Wess-Zumino consistency conditions that vanishes on
$SO(4)$ and so is, {\em ceteris paribus}, just as good a candidate for the
anomalous action. It corresponds to a regularization of the
$SO(5)^2U(1)$ anomaly
such that it is appears entirely in the $U(1)$ symmetry, whereas our
action corresponds to an anomaly that is symmetric with respect to the
broken generators in $U(1)$ and $SO(5)$. Whether there
exist yet more consistent effective actions is an open question. 

The two anomalous effective actions that we have found differ
structurally only in
their higher-order terms. But this does not mean that the
non-uniqueness is phenomenologically inconsequential. Indeed, different
choices of regulator for the anomaly lead to different values of the
coefficient of the leading order term. In particular, the coefficient
that corresponds to an anomaly that is symmetrized amongst all three
currents is $\frac{1}{3}$ that of the coefficient that corresponds to
the anomaly that is contained wholly in a single current. So the
resolution of the non-uniqueness issue will be crucial, if we want to
make inferences about the UV structure of the theory (in particular
its fermionic representation content), using experimental data. 

Even if this non-uniqueness can be resolved, one should also bear in
mind that the couplings of the Goldstone bosons to SM fermions will
also generate loop contributions to the couplings in the anomalous
effective action. 
\subsection*{The WZW term} 
There is a possible WZW term in the model, as
we can see by computing $H^5_{dR} (SO(5) \times
U(1)/SO(4))$.
Recalling that
$SO(n+1)/SO(n)$ and $S^n$ are equivalent
as smooth manifolds, we thus have that 
$H^5_{dR} (SO(5) \times
U(1)/SO(4)) = H^5_{dR} (S^4 \times S^1) = H^4_{dR} (S^4) \otimes H^1_{dR}
(S^1) \simeq \mathbb{R} \otimes \mathbb{R} \simeq \mathbb{R}$ (where
we used the K\"{u}nneth formula and the fact that $H^i_{dR} (S^n)$ vanishes
unless $i=0$ or $i=n$, in which case it is isomorphic to
$\mathbb{R}$). Thus, the
theory admits a WZW term. 

We may easily find the form of the WZW term, at least for field configurations 
that correspond to the trivial class of the
fourth homotopy group. These may be written \cite{D'Hoker:1994ti}  as the integral over a
5-ball, whose boundary is the spacetime $S^4$, of a $G$-invariant 5-form,
whose existence is guaranteed by the the non-vanishing fifth de Rham cohomology group.\footnote{Unfortunately, this trick does not work for a general field
configuration, because the fourth homotopy is $\pi_4 (S^4 \times
S^1) \simeq \pi^4 (S^4) \oplus \pi^4(S^1) \simeq  \mathbb{Z} \oplus 
\{e\} \simeq \mathbb{Z} \neq 0$.} For $G/H \simeq S^4 \times S^1$ it is just the product
of the usual volume forms on the hyperspheres. At leading order in the
fields, we can integrate over the 5-ball to get
\begin{gather}
\int_{S^4} \epsilon_{ijkl}\eta d h^i d h^j d
h^k d h^l,
\end{gather}
where $h^i$ are co-ordinates in the neighbourhood of the identity on
$S^4$.\footnote{This term was also singled out in
  \cite{Franceschini:2015kwy}, but for different reasons.} In $SU(2)\times U(1)$ language, the LO WZW term is $ \eta dH^\dagger \sigma^i
dH dH^\dagger \sigma^i dH$. 

As expected, the leading order term is invariant under the linearly-realized
subgroup $SO(4)$ and changes by a total derivative under a
shift of the Goldstone bosons, corresponding to an infinitesimal
$SO(5)\times U(1)$ transformation.

As we see in Appendix~\ref{app:anlo}, the WZW term does not arise from (\ref{eq:Gamma}), which vanishes when the
gauge fields vanish. Thus, unlike in QCD,
the WZW term and the anomaly are independent, at least for this choice of
regularization of the anomaly.

The WZW term is in fact the leading-order term coupling all 5 Goldstone
bosons to each other. This can be seen by forming lagrangian
invariants of the sigma model in the usual way out of the objects
$d\eta$ and $e^\xi d e^{-\xi}$, which transform as adjoints under $H$. By Lorentz invariance,
all terms involve an even number of derivatives. Terms with no
derivatives are forbidden by the non-linearly realized symmetry, while
terms with two derivatives are forbidden, because such a term must
take the form $\partial_\mu \eta \, \mathrm{tr} \, e^\xi \partial_\mu
e^{-\xi} =0$. A possible term with 4 derivatives
takes the form $\partial_\mu \eta \, \mathrm{tr} \,  (e^\xi \partial_\nu
e^{-\xi}) (e^\xi \partial_\sigma  e^{-\xi}) (e^\xi \partial_\rho
e^{-\xi})$. Since $SO(5)$ is free of triangle anomalies, the trace
term 
must be antisymmetric in its 3 entries and so a non-vanishing
Lorentz-invariant can be obtained only by contracting with
$\epsilon^{\mu \nu \sigma \rho}$, such that we can revert to the
language of differential forms. We have that $e^\xi d
e^{-\xi} = d\xi + \frac{1}{2} [\xi, d\xi] + \dots$, such that the
leading order term involving all Goldstone bosons takes the form
$\frac{3}{2}d\eta \mathrm{tr} d\xi d\xi [\xi,d\xi]$. We need this to
be non-vanishing
when each $\xi$ corresponds to a distinct Goldstone boson and one
easily check using the basis in (\ref{eq:basis}) that this is
not so.

To explore the physics of the WZW term,
we first gauge the SM subgroup. Since this is a subgroup of
$H$, under which the WZW term transforms linearly, we may
follow the
usual prescription of promoting derivatives to covariant derivatives, obtaining $\eta DH^\dagger \sigma^i
DH DH^\dagger \sigma^i DH$. 

Being of high dimension, the WZW leads to small contributions to
low-energy physics. They may, nevertheless, be observable at a future
high-precision collider, if sufficiently exotic. As an example, by the Goldstone boson
equivalence theorem and by the antisymmetry in the fields,
the WZW term leads, after electroweak symmetry breaking, to a coupling involving $\eta, h, W^+,W^-,$ and
$Z$ and hence a possible decay mode $\eta \rightarrow hW^+W^-Z$.

We remark that, whilst the WZW term is the leading order term
coupling all 5 Goldstone bosons to one another, this does not
necessarily imply that it gives the dominant contribution to this
decay mode. Indeed, once we switch on the gauging and other symmetry-breaking couplings, we may
well get contributions to this decay at lower orders, albeit paying
the price of small, symmetry breaking couplings instead.

\subsection*{Discrete symmetries and $Z\rightarrow b
\overline{b}$} \label{app:zbb}
As we have already
remarked, the fact that $SO(5) \times
U(1)/SO(4)$ is a symmetric space means that the Lie algebra possesses
the `internal parity' automorphism $\mathfrak{h} \rightarrow \mathfrak{h}, \mathfrak{k}
\rightarrow -\mathfrak{k}$.  
The terms in the effective action giving rise to production and decay
of the $\eta$ are odd under this, so it could only be a symmetry of
the dynamics if it were accompanied by a spatial inversion. In any
case, the internal parity is broken in the vacuum by the Higgs VEV. 

A more desirable symmetry to have, perhaps, is one that protects the decay rate for $Z\rightarrow b
\overline{b}$ \cite{Agashe:2006at}. In the minimal model based on $G=SO(5)$, this is
achieved by enlarging the linearly-realized subgroup from $SO(4)$ to
$O(4)$.\footnote{In fact, if we wish to include matter fields in the
  theory in spinor representations, then we should consider not
  $SO(5)$ but rather its universal cover $Sp(2)$. As described in
  \cite{Gripaios:2014pqa}, the relevant homogeneous spaces without and with
  custodial protection of $Z\rightarrow b
\overline{b}$ are $Sp(2)/(Sp(1)\times Sp(1))$ and $Sp(2)/(Sp(1)\times
Sp(1) \rtimes \mathbb{Z}_2)$, where the homomorphism in the
semi-direct product maps the non-trivial element in $\mathbb{Z}_2$ to
the outer automorphism of $Sp(1)\times
Sp(1)$ that interchanges the two $Sp(1)$s. The homogeneous spaces are
homeomorphic to $SO(5)/SO(4)$ and $SO(5)/O(4)$, respectively, and the
discussion given here can be carried over straightforwardly.} 
The same enlargement could, of course, be carried out in the model
described here, but it has the consequence that the 
WZW term is forced to vanish. Indeed, the
usual action of $SO(5)$ on $\mathbb{R}^5$ gives rise to transitive
actions on both $S^4$ (included in $\mathbb{R}^5$ as the set of points
equidistant from the origin) and $\mathbb{R}P^4$ (given as the set of
lines through the origin in $\mathbb{R}^5$ and which we may also think
of as the sphere with antipodal points identified). The stability subgroup in
the former case is isomorphic to $SO(4)$, while in the latter case it
is $O(4)$. Thus $SO(5)/SO(4)$ is homeomorphic to $S^4$, while
$SO(5)/O(4)$ is homeomorphic to $\mathbb{R}P^4$. Now, $H^4_{dR}
(\mathbb{R}P^4)$ vanishes,\footnote{The reason for this is that the
  volume form on $S^4$, which is given by the pull-back to $S^4$
via the inclusion map $i:S^4 \rightarrow \mathbb{R}^5$ of the form
$\sum_{i=1}^4 (-1)^i x^i dx^1 \dots dx^4$ (where in the ellipsis we
omit $dx^i$), is not identical at antipodal points; this is consistent
with the non-orientability of $\mathbb{R}P^4$.}
as do its other de-Rham cohomology
groups (excepting of course $H^0_{dR}$), and so the K\"{u}nneth
formula tells us that with $O(4)$ included in this way, $H^5_{dR} (SO(5) \times SO(2)/O(4)) = 0$, such that
there can be no WZW term. 

The WZW term may, however, be resurrected by changing the
inclusion of the custodial $O(4)$ in $G$. To understand this, it is
useful to see more explicitly why the leading-order WZW term is forbidden in the
standard implementation. To this end,
choose co-ordinates $(\vec{h} ,
  1 )$ on the unit 4-sphere included in $\mathbb{R}^5$ in the
neighbourhood of the stability point $( \vec{0} ,
  1 )$. The stability group of the sphere is then
$\{ \begin{pmatrix} O^+ &0 \\ 0& +1
   \end{pmatrix}\} $, where $O^+$ is any 4x4 orthogonal matrix of
   determinant $+1$, and hence is isomorphic to $SO(4)$. But if we identify antipodal points, $( -\vec{h} ,
  -1 )\sim (\vec{h} ,
  1 )$, then the stability subgroup is enhanced to $\{ \begin{pmatrix} O^\pm &0 \\ 0& \pm1
   \end{pmatrix}\} $, where $O^-$ is any 4x4 orthogonal matrix of
   determinant $\pm1$, and hence is indeed isomorphic to $O(4)$, as we
   claimed earlier. Now, under
   the action of an element of $O(4)$ that is disconnected from the identity, $( \vec{h} ,
  1) \rightarrow (O^-\vec{h},
  -1 )\sim (-O^-\vec{h} ,
  +1 )$. Thus the putative leading-order WZW term, which is
proportional to $\epsilon_{ijkl} h^i h^j h^k h^l$ is sent to $(-1)^4
\mathrm{det} O^- = -1$ times itself, and is not invariant under such
transformations. But the leading order WZW term should be invariant under $O(4)$ and
so must vanish. 

Clearly, we can resurrect the WZW term, at least at leading order, by
arranging for the $O(4)$ custodial group to be included in $G$ in such
a way that the action of elements in $O(4)$ disconnected from the
identity also sends $\eta \rightarrow -\eta$. To achieve this, set $G= SO(5) \times O(2)$ and
let $H$ be the subgroup 
\begin{gather}
\{ (\begin{pmatrix} O^+ &0 \\ 0& +1
   \end{pmatrix},\begin{pmatrix} 1 &0 \\ 0& 1
   \end{pmatrix}), (\begin{pmatrix} O^- &0 \\ 0&-1
   \end{pmatrix},\begin{pmatrix} 1 &0 \\ 0& -1
   \end{pmatrix})\}
\end{gather}
 $H$ is still isomorphic to $O(4)$, but now the
   action of elements in $O(4)$ disconnected from the identity sends 
$\eta \rightarrow -\eta$. We conjecture therefore that $H^5_{dR} \neq
0$ in this case, such that there is a WZW term.

\section{Couplings to fermions and flavour violation}
\label{sec:ferm}
We now discuss the couplings of the $\eta$ singlet to SM fermions.
We postulate that the SM fermion Yukawa couplings are generated via the paradigm of Partial Compositeness (PC) \cite{Kaplan:1991dc}. The basic assumption is that elementary states $f^i$ (where $f \in \{Q_L,u_R,d_R,L_L,e_R \}$ and $i$ is the family index) couple linearly to fermionic operators $\overline{\mathcal{O}}^f_i$ of the strong sector:
\begin{equation*}
\mathcal{L}_{\textrm{PC}} = 
g_{\rho} \epsilon^q_i  \, \overline{\mathcal{O}}^{q}_i Q_L^i +
g_{\rho} \epsilon^u_i  \, \overline{\mathcal{O}}^{u}_i u_R^i +
g_{\rho} \epsilon^d_i  \, \overline{\mathcal{O}}^{d}_i d_R^i +
g_{\rho} \epsilon^{\ell}_i  \, \overline{\mathcal{O}}^{\ell}_i L_L^i +
g_{\rho} \epsilon^{e}_i  \, \overline{\mathcal{O}}^{e}_i e_R^i + \textrm{h.c.}
\end{equation*}
We simplify the description of the strong sector as in \cite{Giudice:2007fh}, assuming a single strong coupling $g_{\rho}$, and a single mass scale $m_{\rho}$. The linear mixing parameters $\epsilon^a_i$ are taken to be hierarchical in order to reproduce the pattern of masses and mixing of the SM fermions. 
In particular, it can be shown that the Yukawa couplings of up and
down quarks and of charged leptons are given by
\begin{equation}
Y^U _{ij} \sim g_{\rho} \epsilon^q_i \epsilon^u_j \, , \qquad Y^D_{ij} \sim g_{\rho} \epsilon^q_i \epsilon^d_j \qquad \textrm{and} \qquad Y^E_{ij} \sim g_{\rho} \epsilon^{\ell}_i \epsilon^e_j \, .
\end{equation}
Throughout this Section, we use the symbol $\sim$ to indicate a relation that holds up to an unknown $\mathcal{O}(1)$ complex coefficient whose value is determined by the unknown strong sector dynamics.
As in \cite{KerenZur:2012fr,Gripaios:2014tna}, a viable choice of the mixing parameters is given in Fig.~\ref{fig:yuk}.
\begin{figure}
\begin{center}
\begin{tabular}{c c}
\toprule
Mixing Parameter & Value \\
\midrule
$\epsilon^q_1 = \lambda^3 \epsilon^q_3 $ & $1.15 \times 10^{-2} \, \epsilon^q_3$  \\
$\epsilon^q_2 = \lambda^2 \epsilon^q_3$ & $5.11 \times 10^{-2} \, \epsilon^q_3$ \\
\hline
$\epsilon_1^u= \frac{m_u}{v g_{\rho}} \frac{1}{\lambda^3 \epsilon^q_3}$ & $5.48 \times 10^{-4} / ( g_{\rho} \epsilon^q_3 )$ \\
$\epsilon_2^u= \frac{m_c}{v g_{\rho}} \frac{1}{\lambda^2 \epsilon^q_3}$ & $5.96 \times 10^{-2} / ( g_{\rho} \epsilon^q_3 )$ \\
$\epsilon_3^u= \frac{m_t}{v g_{\rho}} \frac{1}{\epsilon^q_3}$ & 0.866/$( g_{\rho} \epsilon^q_3 )$ \\
\hline
$\epsilon_1^d= \frac{m_d}{v g_{\rho}} \frac{1}{\lambda^3 \epsilon^q_3}$ & $ 1.24 \times 10^{-3} /( g_{\rho} \epsilon^q_3 )$ \\
$\epsilon_2^d= \frac{m_s}{v g_{\rho}} \frac{1}{\lambda^2 \epsilon^q_3}$ & $ 5.29 \times 10^{-3} /( g_{\rho} \epsilon^q_3 )$ \\
$\epsilon_3^d= \frac{m_b}{v g_{\rho}} \frac{1}{\epsilon^q_3}$ & $ 1.40 \times 10^{-2} ( g_{\rho} \epsilon^q_3 )$ \\
\hline
$\epsilon_1^{\ell}= \epsilon_1^{e}=\left(\frac{m_e}{g_{\rho} v}\right)^{1/2}$ & $ 1.67\times 10^{-3} /g^{1/2}_{\rho}$ \\
$\epsilon_2^{\ell}= \epsilon_2^{e}=\left(\frac{m_{\mu}}{g_{\rho} v}\right)^{1/2}$ & $ 2.43 \times 10^{-2} /g^{1/2}_{\rho}$ \\
$\epsilon_3^{\ell}= \epsilon_3^{e}=\left(\frac{m_{\tau}}{g_{\rho} v}\right)^{1/2}$ & 0.101/$g^{1/2}_{\rho}$ \\
\bottomrule
\end{tabular}
\caption{\label{fig:yuk} Partial compositeness mixing parameters and
  values. The input running masses of the SM particles are taken at
  the renormalisation scale of 1 TeV, with $v=174$ GeV.\label{mixing}}
\end{center}
\end{figure}
We remark that we have tacitly assumed, for simplicity, that every elementary
field $f^{a}_i$ couples to a \textit{single} operator of the strong
sector. In that case, it is easy to derive the coupling of the goldstone boson $\eta$
to the fermions $f^i$. Indeed, it is enough to replace $f^i \to f^i \exp
\left(i \frac{\sqrt{2}}{f_{\eta}} \eta \, Z_{f_i} \right)$ in the EFT
of the usual composite Higgs model based on $SO(5) \times U(1)_X$,
where $Z_{f_i}$ is the $U(1)_\eta$ charge.
As we shall see in \S
    \ref{sec:pot}, there is a price to be paid for this assumption,
    namely that one then requires an additional source of explicit
    $U(1)_\eta$ breaking in the model in order to generate a potential for
    the singlet. We expect, however, that relaxing this assumption
    will lead to comparable bounds.

Without specifying the details and quantum numbers of the composite operators under $SO(5)\times U(1)_X$, integrating away the heavy sector at the scale $m_{\rho}$ and keeping the leading term in $H/f$ and $\eta/f_{\eta}$, we get (in complete generality) that
\begin{eqnarray}
\mathcal{L}_{\textrm{yuk}} &=& - Y^{U}_{ij} \, H \overline{Q}^i_L u^j_R \left[ 1 + i \frac{\sqrt{2}}{f_{\eta}} \eta \left(Z_{Q^i_L}-Z_{u_R^j} \right) \right] + \textrm{h.c.}\\
&&- Y^{D}_{ij} \, H^c \overline{Q}^i_L d^j_R \left[ 1 + i \frac{\sqrt{2}}{f_{\eta}} \eta \left(Z_{Q^i_L}-Z_{d^j_R} \right) \right]+ \textrm{h.c.} \\
&&- Y^{E}_{ij} \, H^c \overline{L}^i_L e^j_R \left[ 1 + i \frac{\sqrt{2}}{f_{\eta}} \eta \left(Z_{L^i_L}-Z_{e^j_R} \right) \right] + \textrm{h.c.} 
\end{eqnarray}

The Yukawa couplings are specified in a basis where the SM fields have
specific charge assignments under $U(1)_{\eta}$. The Yukawa matrices are diagonalised by bi-unitary transformations:
\begin{eqnarray}
\hat{Y}^U = L_U Y^U R_U^{\dagger} &=& \frac{1}{v} \, \textrm{diag}(m_u, m_c, m_t) \\ 
\hat{Y}^D = L_D Y^D R_D^{\dagger} &=& \frac{1}{v} \, \textrm{diag}(m_d, m_s, m_b) \\ 
\hat{Y}^E = L_E Y^E R_E^{\dagger} &=& \frac{1}{v} \, \textrm{diag}(m_e, m_{\mu}, m_{\tau}) \, .
\end{eqnarray}
The expected size of the entries of these unitary matrices are linked to the mixing parameters in the following way
\begin{equation}
(L_U)_{ij} \sim (L_D)_{ij} \sim \min \left( \frac{\epsilon^q_i}{\epsilon^q_j}, \frac{\epsilon^q_j}{\epsilon^q_i} \right)  \qquad (R_U)_{ij} \sim \min \left( \frac{\epsilon^u_i}{\epsilon^u_j}, \frac{\epsilon^u_j}{\epsilon^u_i} \right) 
\qquad (R_D)_{ij} \sim \min \left( \frac{\epsilon^d_i}{\epsilon^d_j}, \frac{\epsilon^d_j}{\epsilon^d_i} \right) 
\end{equation}
and similarly for the leptonic sector.

Rewriting the
lagrangian in the mass basis and replacing the Higgs doublet with its
VEV, one may deduce the flavour- and CP-violating couplings of the $\eta$ to SM fermions:
\begin{eqnarray}
\nonumber
\mathcal{L}_{\textrm{yuk}} &\supset& 
-  \sum_{u_i,u_j = u,c,t}  Y_{u_i u_j} \eta \, \bar u_i P_R u_j 
-  \sum_{d,d_j = d,s,b}  Y_{d_i d_j} \eta \, \bar d_i P_R d_j 
- \sum_{\ell_i,\ell_j = e,\mu,\tau}  Y_{\ell_i\ell_j} \eta \, \bar \ell_i P_R \ell_j + \rm h.c.
\end{eqnarray}
The typical size of the induced flavor violating Yukawa couplings depends on the structure dictated by partial compositeness and by the $U(1)_{\eta}$ charge assignment of the different fields.
It is easy to show that
\begin{eqnarray}
Y_{u_i u_j} & = & i \frac{\sqrt{2} \, v}{f_{\eta}} \left[ L_U \hat{Z}_{Q_L} L^{\dagger}_U \hat{Y}_U +   \hat{Y}_U R_U \hat{Z}_{U_R} R^{\dagger}_U\right]_{ij} \\
Y_{d_i d_j} & = & i \frac{\sqrt{2} \, v}{f_{\eta}} \left[ L_D \hat{Z}_{Q_L} L^{\dagger}_D \hat{Y}_D +   \hat{Y}_D R_D \hat{Z}_{D_R} R^{\dagger}_D\right]_{ij} \\
Y_{e_i e_j} & = & i \frac{\sqrt{2} \, v}{f_{\eta}} \left[ L_E \hat{Z}_{L_L} L^{\dagger}_E \hat{Y}_E +   \hat{Y}_E R_E \hat{Z}_{E_R} R^{\dagger}_E\right]_{ij}
\end{eqnarray}
The $\hat{Z}$ matrices are diagonal and contain the $Z$-charges of the fields; in particular we have defined $\hat{Z}_{f} \equiv \textrm{diag}(Z_{f^1}, Z_{f^2},Z_{f^3})$.

With these expressions in hand, let us consider to what extent the suppression provided by the partial compositeness ansatz is sufficient to protect the model from dangerous flavour- and CP-violating contributions to physical observables.

If we assume the `worst-case scenario' of an anarchic charge assignment ($Z_{f^i} = \mathcal{O}(Z)$
for every field $f^i =\{Q_L^i, u_R^i, d_R^i, L^i_L,E^j_R \} $), we
obtain couplings of the following sizes:
\begin{eqnarray}
(Y^U_{\eta})_{ij} \equiv Y_{u_i u_j} \sim g_{\rho} \epsilon^q_i \epsilon^u_j  \frac{\sqrt{2} v}{f_{\eta}} Z &=& 
\frac{\sqrt{2} v}{f_{\eta}} Z
\left(
\begin{array}{ccc}
6.3 \times 10^{-6} & 6.8 \times 10^{-4} & 9.9 \times 10^{-3} \\
2.8 \times 10^{-5} & 3.0 \times 10^{-3} & 4.4 \times 10^{-2} \\
5.4 \times 10^{-4} & 6.0 \times 10^{-2} & 0.87 
\end{array}
\right)\\
(Y^D_{\eta})_{ij} \equiv Y_{d_i d_j} \sim g_{\rho} \epsilon^q_i \epsilon^d_j  \frac{\sqrt{2} v}{f_{\eta}} Z &=& 
\frac{\sqrt{2} v}{f_{\eta}} Z
\left(
\begin{array}{ccc}
1.4 \times 10^{-5} & 6.1 \times 10^{-5} & 1.6 \times 10^{-4} \\
6.3 \times 10^{-5} & 2.7 \times 10^{-4} & 7.1 \times 10^{-4} \\
1.2 \times 10^{-3} & 5.3 \times 10^{-3} & 1.4 \times 10^{-2} 
\end{array}
\right) \\
(Y^E_{\eta})_{ij} \equiv Y_{e_i e_j} \sim g_{\rho} \epsilon^{\ell}_i \epsilon^e_j  \frac{\sqrt{2} v}{f_{\eta}} Z &=& 
\frac{\sqrt{2} v}{f_{\eta}} Z
\left(
\begin{array}{ccc}
2.8 \times 10^{-6} & 4.1 \times 10^{-5} & 1.7 \times 10^{-4} \\
4.1 \times 10^{-5} & 5.9 \times 10^{-4} & 2.5 \times 10^{-3} \\
1.7 \times 10^{-4} & 2.5 \times 10^{-3} & 1.0 \times 10^{-2} 
\end{array}
\right) 
\end{eqnarray}
These couplings are subject to phenomenological constraints. Bounds
derived from flavour and CP violating processes induced by the
exchange of the $\eta$ boson can be found in the model independent
analysis of \cite{Goertz:2015nkp}. We translate these into bounds on the combination $\frac{Z}{f_{\eta}}$, as reported in Fig.~\ref{tab:1}. 
\begin{figure}
\begin{center}
\begin{tabular}{l c c }
\toprule
Bound on $Y_{f,f'}$ & Observable & $Z \left(\frac{f_{\eta}}{700 \textrm{ GeV}} \right)^{-1} \left(\frac{M_{\eta}}{750 \textrm{ GeV}} \right)^{-1}$ \\
\midrule
$\sqrt{{\rm Re}[(Y_{sd})^2]}, \sqrt{{\rm Re}[(Y_{ds})^2]}  < 1.3 \times 10^{-4}  \left(\frac{M_{\eta}}{750 \textrm{ GeV}} \right) $ & $ \Delta m_K$ & $< 5.9 $ \\
$\sqrt{{\rm Re}[(Y_{sd} Y^*_{ds}]}  < 4.6 \times 10^{-5}  \left(\frac{M_{\eta}}{750 \textrm{ GeV}} \right) $ & $ \Delta m_K$ & $< 2.1 $ \\
$\sqrt{{\rm Im}[(Y_{sd})^2]}, \sqrt{{\rm Im}[(Y_{ds})^2]} < 3.4 \times 10^{-6}  \left(\frac{M_{\eta}}{750 \textrm{ GeV}} \right) $ & $\epsilon_K$ & $< 0.15 $ \\
$\sqrt{{\rm Im}[(Y_{sd} Y^*_{ds}]}  < 1.6 \times 10^{-5} \left(\frac{M_{\eta}}{750 \textrm{ GeV}} \right)$ & $\epsilon_K  $ & $< 5.2 \times 10^{-2} $ \\
$\sqrt{{\rm Re}[(Y_{cu})^2]}, \sqrt{{\rm Re}[(Y_{uc})^2]}< 3.3 \times 10^{-4}\left(\frac{M_{\eta}}{750 \textrm{ GeV}} \right)$ & $ x_D $ & $<1.4 $ \\
$\sqrt{{\rm Re}[Y_{cu} Y^*_{uc}]}< 3.9 \times 10^{-5}\left(\frac{M_{\eta}}{750 \textrm{ GeV}} \right)$ & $ x_D $ & $< 0.17 $ \\
$\sqrt{{\rm Im}[(Y_{cu})^2]}, \sqrt{{\rm Im}[(Y_{uc})^2]}< 4.0 \times 10^{-5}\left(\frac{M_{\eta}}{750 \textrm{ GeV}} \right)$ & $ (q/p)_D, \phi_D $ & $< 0.17 $ \\
$\sqrt{{\rm Im}[Y_{cu} Y^*_{uc}]}< 4.0 \times 10^{-5}\left(\frac{M_{\eta}}{750 \textrm{ GeV}} \right)$ & $ (q/p)_D, \phi_D $ & $< 2.0 \times 10^{-2} $ \\
$\sqrt{{\rm Re}[(Y_{bd})^2]}, \sqrt{{\rm Re}[(Y_{bd})^2]}< 4.1 \times 10^{-4}\left(\frac{M_{\eta}}{750 \textrm{ GeV}} \right)$ & $ \Delta m_d $ & $< 7.3 $ \\
$\sqrt{{\rm Re}[Y_{bd} Y^*_{db} ]}, \sqrt{{\rm Re}[(Y_{bd})^2]}<  1.4 \times 10^{-4}\left(\frac{M_{\eta}}{750 \textrm{ GeV}} \right)$ & $ \Delta m_d $ & $< 2.4 $ \\
$\sqrt{{\rm Im}[(Y_{bd})^2]}, \sqrt{{\rm Im}[(Y_{bd})^2]}< 2.3 \times 10^{-4}\left(\frac{M_{\eta}}{750 \textrm{ GeV}} \right)$ & $ \sin2\beta $ & $< 4.1 $ \\
$\sqrt{{\rm Im}[(Y_{bd}) Y^*_{db} ]}< 7.6 \times 10^{-5}\left(\frac{M_{\eta}}{750 \textrm{ GeV}} \right)$ & $ \sin2\beta $ & $< 1.4 $ \\

$|(Y_{bs})|, |(Y_{sb}) | < 1.7\times 10^{-3}\left(\frac{M_{\eta}}{750 \textrm{ GeV}} \right)$ & $ \Delta m_s $ & $<0.91 $ \\
$\sqrt{|Y_{bs} Y^*_{sb}|} < 5.7\times 10^{-4}\left(\frac{M_{\eta}}{750 \textrm{ GeV}} \right)$ & $ \Delta m_s $ & $< 0.31$ \\
\bottomrule
\end{tabular}
\end{center}
\caption{\label{tab:1} Constraints on $\eta$ couplings to SM fermions (first column) derived from low energy precision observables (second column). The limits on $\frac{Z}{f_{\eta} M_{\eta}}$ are presented in the third column. }
\end{figure}
It is clear from these results that, in order to pass the bounds imposed by observables involving the first two families of quarks, we need $Z \left(\frac{f_{\eta}}{700 \textrm{ GeV}} \right)^{-1} \left(\frac{M_{\eta}}{750 \textrm{ GeV}} \right)^{-1} \lesssim 10^{-2}$. The values of $Z$ and $f_{\eta}$ are unknown and depend on the details of the strongly coupled sector. However, the most natural expectation is that  $f_{\eta} \sim f$ and $Z \sim 1$, because the composite Higgs and the composite $\eta$ are generated from the same strong dynamics. If this is the case, an extra source of flavour protection is required.
An easy fix to this problem is to assume that the $\eta$ PNGB couples
to flavour in a universal way. More specifically, we can impose that
$Z_{f^i} = Z_f$ for $i=\{ 1,2,3 \}$. In this case the $\eta$ and the
Higgs boson couplings to fermions are aligned in each sector, such that
\begin{eqnarray}
(Y^{U}_{\eta})_{ij} &=& i \delta_{ij} \frac{m_i^U}{f_{\eta}} \left(Z_{Q_L} - Z_{u_R}\right) \qquad m_i^U = \{m_u,m_c,m_t \} \\
(Y^{D}_{\eta})_{ij} &=& i \delta_{ij} \frac{m_i^D}{f_{\eta}} \left(Z_{Q_L} - Z_{d_R}\right) \qquad m_i^D = \{m_d,m_s,m_b \} \\
(Y^{E}_{\eta})_{ij} &=& i \delta_{ij} \frac{m_i^E}{f_{\eta}} \left(Z_{L_L} - Z_{e_R}\right) \qquad m_i^E = \{m_e,m_{\mu},m_{\tau} \} 
\end{eqnarray}
All the flavour and CP problems are solved, since this pattern is
flavour diagonal.\footnote{There remain, however, sub-dominant flavour
  violating contributions from possible derivative operators, analogous to
  those described in \cite{Agashe:2009di}.} It is, moreover, rather predictive. Indeed the
$\eta$, like the Higgs, couples predominantly to the third
generation. This could have important implications for the production
and decay mechanisms of the singlet, as we now discuss. 

In the narrow width approximation the prompt $\eta$ production at the LHC can be expressed in terms of the relevant decay widths 
\begin{equation}
\sigma (pp \to \eta) = \frac{1}{M_{\eta} \, s} \sum_{P} C_{P\overline{P}} (M_{\eta},s) \, \Gamma_{P\overline{P}} \, ,
\end{equation}
where $\sqrt{s}$ is the center of mass energy of the collider and  $ C_{P\overline{P}} (M_{\eta},s) $ parametrise the relevant parton luminosities. In our framework the relevant partons to be taken into account are expected to be the gluons (if the associated anomalous term in Eq.(\ref{eq:5d}) is present) and the bottom quarks. 
The explicit expressions for the partial widths are given by
\begin{eqnarray}
\Gamma (\eta \to gg) & = & c_3^2 \frac{\alpha_s^2}{8 \pi^3} \frac{M^3_{\eta}}{f^2_{\eta}} , \\
\Gamma (\eta \to b \bar{b}) & = & \frac{3 M_{\eta}}{8 \pi} (Y^D_{\eta})^2_{33} .
\end{eqnarray}
The mechanism of partial compositeness allows also to predict the dominant decay mode to be into top-quarks if $m_{\eta} > 2 \, m_t$. Depending on the value of the mass of the PNGB, phase space could be important and the expression for the decay width in this channel is given by
\begin{eqnarray}
\Gamma (\eta \to t \bar{t}) & = & \frac{3 M_{\eta}}{8 \pi} (Y^U_{\eta})^2_{33} \left(1-\frac{4 m_t^2}{m^2_{\eta}} \right)^{3/2}.
\end{eqnarray}
The large coupling of $\eta$ with the heaviest fermion allows for its production at LHC in association  with top quarks. A recent analysis from the ATLAS collaboration \cite{ATLASTT}, using data at $\sqrt{s}= 13 $ TeV, leads to the following bound:
\begin{equation}
\sigma ( pp \to \eta + \overline{t} t) \times Br (\eta \to \overline{t} t) \lesssim \mathcal{O}(10^{-1}) \textrm{ pb}
\end{equation}
for a mass of the PNGB $m_{\eta} < 1 $ TeV.

We conclude this section noticing that the simple flavour structure that we have just described, while guaranteeing immunity from  flavour problems, does not allow one to generate a scalar potential
 (and hence a mass) for the singlet from fermionic couplings. As we
 discuss in the next Section, to do so requires that at least one of the elementary fermions in the partial compositeness
  scenario mixes with multiple strong-sector operators. Even if one
  tries to do so in a way that is as safe as possible (for example by allowing
 the right-handed up quarks to couple to
strong-sector operators with just two values of the $U(1)_\eta$ charge), one ends
up re-introducing flavour-violation in the right-handed up sector
at a level comparable to that obtained with anarchic charge
assignments in Fig.~\ref{tab:1}, which is itself comparable to that
obtained in the minimal composite Higgs model. Thus, if one wishes to
generate the scalar potential from fermionic couplings, it would seem
that either a mild tuning or some kind of flavour-alignment mechanism (such as those advanced
in \cite{Redi:2011zi}) is required.
\section{The scalar potential}
\label{sec:pot}

Since the $\eta$ singlet is protected by a shift symmetry, its mass
and non-derivative interactions must be proportional to
$U(1)_\eta$-breaking couplings. The elementary fermion couplings to
the strong sector are the main source of such global symmetry
violations, and the $\eta$ singlet then obtains a potential via the
same Coleman-Weinberg mechanism that radiatively generates the
pseudo-Goldstone Higgs potential at one loop. This must originate from
fermion couplings, since no potential is generated by gauge couplings
in the absence of anomalies,
because $U(1)_\eta$ commutes with the rest of $G$.\footnote{In the
  presence of anomalies and without other sources of
  $U(1)_\eta$-breaking, the $\eta$ plays the role of an electroweak
  axion. The resulting contributions to its mass are thus completely negligible
  compared to those considered here.} The particular form of the symmetry breaking from Yukawa
couplings is, in general, model-dependent. 

To illustrate the mechanism in a minimal phenomenological model, we
take an elementary top-right coupling to two strong-sector operators
with different $U(1)_\eta$ charges such that the symmetry is
explicitly broken by a collective mechanism.\footnote{This is in
  contrast to various composite Higgs models where the top right is a
  fully composite state.} The doubling of the top-right operator
  is necessary to break the $U(1)_\eta$ symmetry, since with only one
  operator, we can restore it by assigning a suitable
  $U(1)_\eta$ charge
  to the elementary fermion.

The simplest realisation of the model is to extend the minimal composite Higgs with the elementary fermions $q_L, u_R,$ and $d_R$ uplifted to a spinorial representation of $SO(5)$, under which the corresponding composite operators $\mathcal{O}_q, \mathcal{O}_{u_1}, \mathcal{O}_{u_2}$ and $\mathcal{O}_{d}$ transform. Summing implicitly over three flavours, the relevant Lagrangian terms may be written as
\begin{equation}
\mathcal{L} \supset g_\rho \epsilon^q \overline{q}_L\mathcal{O}_q + g_\rho\epsilon^{u_1} \overline{u}_R \mathcal{O}_{u_1} + g_\rho\epsilon^{u_2} \overline{u}_R \mathcal{O}_{u_2} + g_\rho\epsilon^{d}\overline{d}_R\mathcal{O}_{d} + \text{h.c.} \quad .
\end{equation}
We see that if one of the two top-right couplings is set to zero then a $U(1)_\eta$ symmetry may be restored. The doubling of the corresponding  $u_R$ operator thereby provides a collective mechanism for breaking the symmetry. The elementary fermions embedded in complete spinorial representations of $SO(5)$ decompose as ${\bf 4} = ({\bf 2}, {\bf 1}) + ({\bf 1}, {\bf 2})$ under $SU(2)_L \times SU(2)_R$. By completing the representation with spurious fermions they can be represented by fields transforming under this symmetry as
\begin{equation*}
\Psi_q = \column{ q_L \\ 0 }^{Z_q}_\frac{1}{6} \, , \, \Psi_{u_1} = \column{0 \\ \column{u_R \\ 0} }^{Z_{u_1}}_\frac{1}{6} \, , \, \Psi_{u_2} = \column{0 \\ \column{u_R \\ 0} }^{Z_{u_2}}_\frac{1}{6} \, , \, \Psi_d = \column{0 \\ \column{0 \\ d_R}}^{Z_{d}}_\frac{1}{6} \, ,
\end{equation*}
where the superscript $Z$ represents the $U(1)_\eta$ charges and the subscript is the $U(1)_X$ charge assigned by requiring $Y = T^3_R + X$. We have set to zero the non-dynamical spurions that complete the $SU(2)_L$ ($SU(2)_R$) representation in the upper (lower) two components of the multiplet, though they are formally required to restore the global $SO(5)$ symmetry. 

The Coleman-Weinberg effective potential may be derived by writing the most general $SO(5)\times U(1)_X \times U(1)_\eta$-invariant effective action then setting the spurions to zero to recover the effective Lagrangian, as detailed in Appendix~\ref{app:potential}. The quadratic terms in the background of the Higgs and singlet are then responsible for the one-loop effective action. Assuming real CP-conserving form factors, we obtain for the third-generation $q_L = (t_L, b_L), t_R$ sector, in momentum space, 
\begin{align}
\mathcal{L} &= \overline{q}_L\slashed{p}\left[\Pi_0^q(p) + \Pi_1^q(p) c_h\right]q_L  \nonumber \\
&+ \overline{t}_R\slashed{p}\left[ \Pi_0^{12}(p)+\Pi_0^{u_{12}}(p)c^{12}_\eta - \left(\Pi_1^{12}(p) + \Pi_1^{u_{12}}(p)c^{12}_\eta\right)c_h  \right]t_R + \text{h.c.} \nonumber \\
&+ \overline{q}_L\left[ M_1^{u_1}(p)U_{q1} + M_1^{u_2}(p)U_{q2}\right]s_h H^c t_R + \text{h.c.} \, ,
\label{eq:effectiveaction}
\end{align}
where $H^c \equiv i\sigma^2 H$ with $H$ the usual complex Higgs doublet and
\begin{equation*}
U_{rs} \equiv e^{i\frac{\sqrt{2}}{f_\eta}\left(Z_r - Z_s\right)\eta} = c^{rs}_\eta + is^{rs}_\eta  \, .
\end{equation*}
We have also defined $c_h \equiv \cos{\left(h/f\right)}$, $s_h \equiv \sin{\left(h/f\right)}$, $c_\eta^{rs} \equiv \cos{\left(\sqrt{2}(Z_r-Z_s)\eta/f_\eta\right)}$, and $s_\eta^{rs} \equiv \sin{\left(\sqrt{2}(Z_r-Z_s)\eta/f_\eta\right)}$, with $h \equiv \sqrt{h^a h^a}$, $a = 1,2,3,4$. The $\Pi_{0,1}, M_{1}$ functions are form factors that encapsulate effects from strong dynamics. The resulting potential is detailed in Appendix~\ref{app:potential} with the leading-order approximation found to be of the form 
\begin{align}
V(h,\eta) &\simeq \left(\alpha + \alpha_{12}c_\eta^{12} \right) c_h -
            \left( \beta + \beta_{12} c_\eta^{12} \right)s_h^2 \, , 
\end{align}
where $\alpha, \beta, \alpha_{12},$ and $\beta_{12}$ are coefficients
related to momentum integrals of the $\pi_{0,1}, M_{1}$ form factors. 
Thus, the resulting potential is almost identical to that
  obtained in the minimal model in \cite{Agashe:2004rs}, but with the
  coefficients replaced by $\alpha, \beta \rightarrow \alpha +
  \alpha_{12}c_\eta^{12}, \beta + \beta_{12} c_\eta^{12}$.

The potential has extrema occuring at $s_\eta^{12} = 0 \implies
c_\eta^{12} = \pm 1$\footnote{We remark that a vacuum with
  $c_\eta^{12} = - 1$ does not imply spontaneous violation of $CP$,
  because $CP$ sends $\eta \rightarrow -\eta$ and because physics
  is periodic in the argument of the cosine.} and $c_h = - \frac{1}{2} \frac{\alpha \pm
  \alpha_{12}}{\beta \pm \beta_{12}}$. As is usual in composite Higgs
models, we find that with $O(1)$ values for the coefficients, $v \sim
f$ is expected and so a slight tuning is needed to obtain the required
suppression of the the weak scale for compatibility with electroweak
precision tests. 

There is no mixing between the
Higgs and $\eta$, so no risk of running into bounds from existing
observations in the Higgs sector. The non-vanishing second derivatives
are given by
\begin{align}
\frac{\partial V}{ \partial \eta^2} &= \mp \frac{1}{f_\eta^2} (\alpha_{12} c_h +
                                      \beta_{12} s_h^2)\\
\frac{\partial V}{ \partial h^2} &= -\frac{1}{f^2} \left[ (\alpha \pm \alpha_{12}) c_h -2
                                   (\beta \pm \beta_{12})c_{2h}
                                   \right] =
                                   \frac{2}{f^2}(\beta \pm \beta_{12})
                                   s_h^2 = \frac{2}{f^2}\frac{v^2}{f^2}  (\beta \pm \beta_{12})
\end{align}
Thus, once we have tuned the
electroweak vev to be small compared to $f$, we will also obtain a
corresponding suppression
of the Higgs mass-squared, exactly as one finds in \cite{Agashe:2004rs}. The mass of $\eta$, however, is unsuppressed, so we 
obtain a hierarchy of scales, of parametric size $v/f$ (assuming $f_\eta \sim f$) between the $\eta$ mass
and either the electroweak scale or the mass of the Higgs boson.

An identical conclusion is reached if we instead embed the elementary
fermions in the fundamental, 5-d representation of $SO(5)$, as we describe
in Appendix~\ref{app:potential}. (In this case, as we discuss in
Appendix~\ref{app:zbb}, we can also protect the
$Zb_L\overline{b}_L$ coupling by a custodial symmetry.)
The hierarchy of scales is, in fact, generic, and follows from the fact that the
scalar potential is an even function of $h$. Indeed, with $V(h,\eta) =
f(h^2, \eta)$, we obtain that ${\partial V}{\partial h^2} = 4v^2
{\partial f}/{\partial h^2}|_v$, at an electroweak-symmetry-breaking
minimum. One can also see that any mixing between the mass eigenstates
of the singlet and the Higgs will also be $v/f$ suppressed.

Although this hierarchy of scales is generic, it may be affected by
the well-known difficulty (see {\em e.g.} \cite{Panico:2012uw} for a
comprehensive discussion) of accommodating a Higgs mass as low as 125
GeV in composite Higgs models, given the size
of contributions to the Higgs potential from top quark loops. If the required
additional suppression is an accidental tuning, then we expect no
corresponding suppression in the $\eta$ mass. But if it is achieved by
the presence of light top partners that cut off all contributions to
the scalar potential, then one should find a corresponding suppression
of the $\eta$ mass.

\section{Conclusion}
\label{sec:conclusion}

Composite Higgs models remain viable possibilities for solving the electroweak hierarchy problem. Here we introduced the most minimal extension of the coset structure allowing a non-trivial anomaly structure and discussed the details of the low-energy action reproducing
the anomalies. We showed that there can be higher-order corrections,
beyond dimension 5, to the action reproducing the $SO(5)^2 U(1)$
anomaly, but also pointed out that the effective action is not
unique. We also showed that the structure of the coset
space admits a possible Wess-Zumino-Witten term, by which we mean a
term in the effective lagrangian which is not invariant under the
non-linearly realized symmetries, but rather shifts by a total
derivative. Unlike in QCD, this term is not contained in the anomalous effective action that we
consider. If present, the term leads to an exotic phenomenological
signature in the form of the singlet decay $\eta \rightarrow h W^+ W^- Z$. 

The discussion of the anomaly structure in this specific model
highlights three questions that it would be interesting to resolve in
models based on a general coset space, $G/H$. Firstly: is there a way
to resolve the non-uniqueness issue of the low-energy anomalous
effective action? Secondly: do
Wess-Zumino-Witten terms that are not required to reproduce triangle
anomalies have some other purpose? Thirdly, is there an elegant way to write the
Wess-Zumino-Witten term for coset spaces whose fourth homotopy group
is non-vanishing? 

The anomaly-induced production and decays of the singlet may induce flavour
violation of its couplings to fermions and we have shown how they can be kept under control without fine-tuning if the $\eta$ couples in a 
flavour-universal way through the mechanism of partial compositeness.
For natural $\mathcal{O}(1)$ charge assignments, this pattern of
coupling 
predicts a large decay width through the $t\overline{t}$ final state.

We also showed how the potential for the PNGB Higgs and singlet can be
 generated by elementary fermion couplings to the strong sector that
 break the global symmetry, though this requires a slight departure
   from the flavour-universal pattern of couplings, because of the need
   for a collective breaking mechanism to give mass to the singlet. We
   find that the singlet mass is naturally unsuppressed relative to the
   Higgs mass and electroweak scale, thus requiring no additional
   tuning beyond the usual ones needed for a small electroweak scale
   and light Higgs mass in composite models. Since the form of the potential contains no mixing between the Higgs and the singlet there are no further bounds from the Higgs sector. 

Should the Higgs arise as a pseudo-Nambu-Goldstone boson, it will be imperative to determine the new strong sector responsible for it. Given our current limited understanding of strongly-coupled
theories, the anomaly structure, if present, may be crucial in gaining
some insight as to the
nature of the underlying UV dynamics. We hope that the model described
here, or some variant thereof, may be useful in this regard.

\section*{Acknowledgements}
BG acknowledges
the support of the Science and Technology Facilities Council (grant
ST/L000385/1) and King's College,
Cambridge and thanks O.~Randal-Williams for discussions.
TY is supported by a Research Fellowship from Gonville and Caius College, Cambridge.

\appendix
\section{Scalar potential computations}
\label{app:potential}

\subsection*{Higgs-singlet potential in the extension of the MCHM4}

The elementary fermions are uplifted to a {\bf 4} of $SO(5)$, which decomposes as ${\bf 4} = ({\bf 2}, {\bf 1}) + ({\bf 1}, {\bf 2})$ under $SU(2)_L \times SU(2)_R$. The most general $SO(5) \times U(1)_X \times U(1)_\eta$-invariant effective action up to quadratic order can then be written in momentum space as
\begin{align}
\mathcal{L} &= \sum_{r = q, u_1, u_2, d} \overline{\Psi}_r \slashed{p}\left[\Pi^r_0(p) + \Pi^r_1(p)\Gamma^i\Sigma_i\right]\Psi_r
+ \left\{\overline{\Psi}_{u_1}\slashed{p}\left[\Pi_0^{u_{12}}(p) + \Pi_1^{u_{12}}(p)\Gamma^i\Sigma_i\right]U_{12} \Psi_{u_2} + \text{h.c.} \right\} \nonumber \\
& \quad\quad\quad\quad +  \sum_{r=u_1,u_2,d}\left\{ \overline{\Psi}_q \left[M_0^r(p) + M_1^r(p)\Gamma^i\Sigma_i\right]U_{qr}\Psi_r + \text{h.c.} \right\} \, ,
\end{align}
where the pseudo-Goldstone singlet $\eta$ and Higgs doublet $h^a = (h^1, h^2, h^3, h^4)$ are given here by 
\begin{align}
\Sigma^i &= \frac{s_h}{h}\left( h^1, h^2, h^3, h^4, h\frac{c_h}{s_h}\right) \nonumber \\
U_{rs} &= e^{i\frac{\sqrt{2}}{f_\eta}\left(Z_r - Z_s\right)\eta} = c^{rs}_\eta + is^{rs}_\eta \, ,
\end{align}
We recall the definitions $h \equiv \sqrt{h^ah^a}$, $c_h \equiv \cos{\left(h/f\right)}$, $s_h \equiv \sin{\left(h/f\right)}$, $c_\eta^{rs} \equiv \cos{\left(\sqrt{2}(Z_r-Z_s)\eta/f_\eta\right)}$, and $s_\eta^{rs} \equiv \sin{\left(\sqrt{2}(Z_r-Z_s)\eta/f_\eta\right)}$. Explicit expressions for the $SO(5)$ gamma matrices $\Gamma^i$ can be found in Ref.~\cite{Agashe:2004rs}. The $\Pi(p), M(p)$ functions are form factors that encapsulate information from the strong sector. 

Setting to zero the non-dynamical spurions that complete the $\Psi$ representation, we obtain the quadratic terms in the Lagrangian for the third generation $q_L = (t_L, b_L), t_R$ sector, 
\begin{align}
\mathcal{L} &= \overline{q}_L\slashed{p}\left[\Pi_0^q(p) + \Pi_1^q(p) c_h\right]q_L 
+ \overline{t}_R\slashed{p}\left[\Pi_0^{u_1} + \Pi_0^{u_2} - \left(\Pi_1^{u_1} + \Pi_1^{u_2}\right)c_h\right]t_R \nonumber \\
&+ \overline{t}_R\slashed{p}\left[\Pi_0^{u_{12}} - \Pi_1^{u_{12}} c_h \right]U_{Z_1 - Z_2} t_R + \text{h.c.} \nonumber \\
&+ \overline{q}_L\left(M_1^{u_1}U_{Z_q - Z_q} + M_1^{u_2} U_{Z_q - Z_2}\right)s_h \hat{H}^c t_R + \text{h.c.} \, ,
\end{align}
where $H^c \equiv i\sigma^2 H$ and $H$ is the complex Higgs doublet. Assuming real CP-conserving form factors, this becomes
\begin{align}
\mathcal{L} &= \overline{q}_L\slashed{p}\left[\Pi_0^q(p) + \Pi_1^q(p) c_h\right]q_L  \nonumber \\
&+ \overline{t}_R\slashed{p}\left[ \Pi_0^{12}(p)+\Pi_0^{u_{12}}(p)c^{12}_\eta - \left(\Pi_1^{12}(p) + \Pi_1^{u_{12}}(p)c^{12}_\eta\right)c_h  \right]t_R + \text{h.c.} \nonumber \\
&+ \overline{q}_L\left[ M_1^{u_1}(p)U_{q1} + M_1^{u_2}(p)U_{q2}\right]s_h H^c t_R + \text{h.c.} \, ,
\end{align}
where $\Pi_{0,1}^{12} \equiv \Pi_{0,1}^{u_1} + \Pi_{0,1}^{u_2}$. Including the $SU(2)_L$ gauge field contributions with form factors $\Pi_0$ and $\Pi_1$ as defined in Ref.~\cite{Agashe:2004rs}, the resulting Coleman-Weinberg potential generated at one loop is given by 
\begin{align}
V(h,\eta) &= -2N_c \int\frac{d^4p}{(2\pi)^4} \left\{2\log\left(1 + \frac{\Pi_1^q}{\Pi_0^q}c_h\right) + \log\left(1 - \frac{\Pi_1^{12} + \Pi_1^{u_{12}}c_\eta^{12} }{\Pi_0^{12} + \Pi_0^{u_{12}}c_\eta^{12}}c_h\right) \right. \nonumber \\
 &\left. + \log\left(1 - \frac{|M^{u_1}_1U_{q1} + M^{u_2}_1U_{q2}|^2 }{p^2\left(\Pi_0^q + \Pi_1^q c_h\right)\left[\left(\Pi_0^{12} - \Pi_1^{12}c_h\right) + \left(\Pi_0^{u_{12}} - \Pi_1^{u_{12}}c_h\right)c^{12}_\eta \right]}s_h^2 \right)  \right\} \nonumber \\
 &+ \frac{9}{2}\int\frac{d^4p}{(2\pi)^4}\log\left(1 + \frac{1}{4}\frac{\Pi_1}{\Pi_0}s_h^2\right) \, .
\end{align}
Assuming the form factors decrease fast enough with increasing momentum, the logarithm may be expanded to give the leading-order approximation for the potential,
\begin{align}
V(h,\eta) &\simeq \left(\alpha + \alpha_{12}c_\eta^{12}\right)c_h - \left(\beta + \beta_{12}c_\eta^{12}\right)s_h^2 \, .
\end{align}
The coefficients are related to the form factor integrals as
\begin{align*}
\alpha &= 2N_c \int \frac{d^4 p}{(2\pi)^4}\left(\frac{\Pi_1^{12}}{\Pi_0^{12} + \Pi_0^{u_{12}}} - 2\frac{\Pi_1^q}{\Pi_0^q} \right) \quad , \quad 
\alpha_{12} = 2N_c\int \frac{d^4 p}{(2\pi)^4}\left(\frac{\Pi_1^{u_{12}}}{\Pi_0^{12} + \Pi_0^{u_{12}}} \right) \, , \\
\beta_V &= - \int \frac{d^4 p}{(2\pi)^4}\frac{9}{8}\frac{\Pi_1}{\Pi_0} \quad , \quad 
\beta_{1,2} = 2N_c\int \frac{d^4 p}{(2\pi)^4}\left( \frac{\left(M_1^{u_{1,2}}\right)^2}{\left(-p^2\right)\left(\Pi_0^q + \Pi_1^q\right)\left(\Pi_0^{12} + \Pi_0^{u_{12}} - \Pi_1^{12} - \Pi_1^{u_{12}}\right)} \right) \, , \\
\beta_{12} &= 2N_c \int \frac{d^4 p}{(2\pi)^4}\left(\frac{2M_1^{u_1}M_1^{u_2}}{\left(-p^2\right)\left(\Pi_0^q + \Pi_1^q\right)\left(\Pi_0^{12} + \Pi_0^{u_{12}} - \Pi_1^{12} - \Pi_1^{u_{12}}\right)} \right) \quad ,
\end{align*}
with $\beta \equiv \beta_V + \beta_1 + \beta_2$.

\subsection*{Higgs-singlet potential in the extension of the MCHM5}
The elementary fermions may instead be embedded in the fundamental representation of $SO(5)$. Such a setup can also be extended to protect the $Zb_L\overline{b}_L$ coupling by a custodial symmetry if we assume that $q_L$ is embedded such that it couples to two  operators with different $U(1)_X$ charges. The resulting Lagrangian of the effective coupling to the composite operators can be written as
\begin{equation*}
\mathcal{L} = g_\rho \epsilon^{q_1} \overline{q}_L\mathcal{O}_{q_1} + g_\rho \epsilon^{q_2} \overline{q}_L\mathcal{O}_{q_2} + g_\rho\epsilon^{u_1}\overline{u}_R\mathcal{O}_{u_1} + g_\rho\epsilon^{u_2}\overline{u}_R\mathcal{O}_{u_2} + g_\rho\epsilon^d \overline{d}_R\mathcal{O}_d + \text{h.c.} \, .
\end{equation*}
The fields transforming under the {\bf 5} of $SO(5)$ with non-dynamical spurions completing the representation (which we again set here to zero) are chosen to be
\begin{small}
\begin{align*}
\Psi_{1_L} = \frac{1}{\sqrt{2}}\column{-b_L \\ ib_L \\ t_L \\ -it_L \\ 0}^{Z_{q_1}}_\frac{2}{3} \, , \, 
\Psi_{2_L} = \frac{1}{\sqrt{2}}\column{ t_L \\ -it_L \\ b_L \\ -ib_L \\ 0}^{Z_{q_2}}_{-\frac{1}{3}} \, , \, 
\Psi_{{1,2}_R} = \column{0 \\ 0 \\ 0 \\ 0 \\ u_R}^{Z_{1,2}}_\frac{2}{3} \, , \, 
\Psi_{d_R} = \column{0 \\ 0 \\ 0 \\ 0 \\ d_R}^{Z_d}_{-\frac{2}{3}} \, .
\end{align*}
\end{small}
The superscripts and subscripts denote the $U(1)_\eta$ and $U(1)_X$ charges respectively. It might initially seem that an explicit breaking of $U(1)_\eta$ from $q_L$ coupling to two different operators will generate a potential for the singlet, thus making the doubling of the top-right couplings redundant, but it turns out that the unbroken $U(1)_X$ symmetry forbids the necessary $\eta$ coupling in the effective action. For this reason we minimally extend the top-right sector as in the previous model and fix  $Z_{q_1} = Z_{q_2} = Z_q$.

The most general effective action under $SO(5)\times U(1)_X\times U(1)_\eta$ is then
\begin{align}
\mathcal{L} &= \sum_{r=1,2} \overline{\Psi}^i_{r_L}\slashed{p}\left(\delta^{ij}\hat{\Pi}^{r_L}_0 + \Sigma^i\Sigma^j\hat{\Pi}_1^{r_L}\right)\Psi^j_{r_L} 
+ \sum_{r=1,2,d}\overline{\psi}_{r_R}^i \slashed{p}\left(\delta^{ij}\hat{\Pi}_0^{r_R} + \Sigma^i\Sigma^j\hat{\Pi}_1^{r_R}\right)\Psi^j_{r_R} \nonumber \\ 
& + \left[ \sum_{r=1,2}\overline{\Psi}^i_{1_L}\left(\delta^{ij}\hat{M}_0^{1rL} + \Sigma_i\Sigma^j\hat{M}_1^{1rL}\right)U_{qr} \Psi_{r_R}^j 
+ \overline{\Psi}^i_{2_L}\left(\delta^{ij}\hat{M}_0^{2bL} + \Sigma^i\Sigma^j\hat{M}_1^{2bL}\right)U_{qb}\Psi^j_{d_R} \right. \nonumber \\ 
&\left. \quad\quad + \overline{\Psi}^i_{1_R}\slashed{p}\left(\delta^{ij}\hat{\Pi}_0^{12R} + \Sigma^i\Sigma^j \hat{\Pi}_1^{12R}\right)U_{12} \Psi^j_{2_R} + \text{h.c.} \vphantom{\sum_x} \right] \, .
\end{align}
Setting the non-dynamical spurions to zero to keep the relevant terms for computing the Coleman-Weinberg effective potential, omitting the bottom contributions, we find
\begin{align}
\mathcal{L} &= \overline{q}_L\slashed{p}\left[\Pi_0^q + \frac{1}{2}s_h^2\left(\Pi_1^{q_1}\hat{H}^c \hat{H^c}^\dagger + \Pi_1^{q_2}\hat{H}\hat{H}^\dagger\right)\right]q_L + \overline{u}_R\slashed{p}\left[\Pi_0^u + \frac{1}{2} s_h^2 \Pi_1^u\right] u_R \nonumber \\
&+ \left[ \frac{1}{\sqrt{2}}c_h s_h \left(M_1^{qu1}U_{q1} + M_1^{qu2}U_{q2}\right)\overline{q}_L\hat{H}^c u_R + \text{h.c.} \right] \nonumber \\
&+ \overline{u}_R\slashed{p}\text{Re}\left\{\left(\Pi_0^{uu} + \frac{1}{2}s_h^2\Pi_1^{uu}\right)U_{12}\right\}u_R \, ,
\end{align}
where
\begin{align}
&\Pi_0^q \equiv \hat{\Pi}_0^{1L} + \hat{\Pi}_0^{2L} \quad , \quad \Pi_1^{q1, q2} \equiv \hat{\Pi}_1^{1L,2L}  \quad ,  \quad \Pi_1^u \equiv -2\left(\hat{\Pi}^{1R}_1 + \hat{\Pi}_1^{2R}\right) \quad , \quad \Pi_1^{uu} \equiv -2\hat{\Pi}_1^{12R} \, , \nonumber \\
&\Pi_0^u \equiv \hat{\Pi}_0^{1R} + \hat{\Pi}_0^{2R} + \hat{\Pi}_1^{1R} + \hat{\Pi}_1^{2R} \quad , \quad 
M_1^{qu1, qu2} \equiv \hat{M}_1^{11L, 12L} \quad , \quad \Pi_0^{uu} \equiv \hat{\Pi}_0^{12R} + \hat{\Pi}_1^{12R} \, .
\end{align}
Assuming real form factors with CP conservation, in the unitary gauge this gives for the top quark sector the quadratic Lagrangian
\begin{align}
\mathcal{L} &= \overline{t}_L \slashed{p} \left[ \Pi_0^q + \frac{1}{2}s_h^2\Pi_1^{q1}\right] t_L + \overline{t}_R\slashed{p}\left[\Pi_0^u + \Pi_0^{uu} + \frac{1}{2}s_h^2\left(\Pi_1^u + \Pi_1^{uu} c_\eta^{12}\right)\right]t_R \nonumber \\ 
&+ \left[ \frac{1}{\sqrt{2}}\left(M_1^{qu1}U_{q1} + M_1^{qu2}U_{q2}\right)c_h s_h \overline{t}_L t_R + \text{h.c.} \right] \, .
\end{align}
The resulting Coleman-Weinberg potential is
\begin{align}
V(h,\eta) &= \frac{9}{2}\int\frac{d^4 p}{(2\pi)^4}\log\left(1 + \frac{1}{4}\frac{\Pi_1}{\Pi_0}s^2_h\right) 
-2N_c \int\frac{d^4 p}{(2\pi)^4}\left\{ \vphantom{\left|\frac{\left|X^{xx}\right|^2}{X^{xx}}\right|^X } \log\left(1 + \frac{1}{2}\frac{\Pi_1^{q1}}{\Pi_0^q}s_h^2\right) \right. \nonumber \\
& \left. + \log\left(1 + \frac{1}{2}\frac{\Pi_1^{q_2}}{\Pi_0^q}s_h^2\right) 
+ \log\left(1 + \frac{1}{2}\frac{\left(\Pi_1^u + \Pi_1^{uu} c_\eta^{12}\right)s_h^2}{\Pi_0^u + \Pi_0^{uu}}\right) 
\right. \nonumber \\
& \left. + \log\left(1 - \frac{\frac{1}{2}\left|M_1^{qu1}U_{q1} + M_1^{qu2}U_{q2}\right|^2c_h^2 s_h^2}{p^2\left[\Pi_0^u + \Pi_0^{uu} + \frac{1}{2}s_h^2\left(\Pi_1^u + \Pi_1^{uu}c_\eta^{12}\right)\right]\left[\Pi_0^q + \frac{1}{2}s_h^2\Pi_1^{q1}\right]} \right)  \right\} \, ,
\end{align}
which may be simplified to the form  
\begin{align}
V(h,\eta) &\simeq \left(\alpha + \alpha_{12} c^{12}_\eta  \right)s^2_h - \left(\beta + \beta_{12} c^{12}_\eta  \right) c^2_h s^2_h \, . 
\end{align}

\section{Higher-order contributions to the anomalous effective action}
\label{app:anlo}
To compute higher-order contributions to the anomalous
effective action (\ref{eq:Gamma}) for the $SO(5)\times U(1)/SO(4)$ model, it is useful to consider
what happens if we start from some $G/H$ and add an additional,
broken, ungauged, $U(1)$ factor, along with a $G^2 U(1)$ triangle anomaly. We thus need to add a Goldstone boson
$\eta$ to the existing Goldstone bosons, $\xi$, and to make the
replacements $A_t \rightarrow A_t - td\eta \implies  F_t \rightarrow
F_t,A_t^2\rightarrow A_t^2$.

We observe that $\eta$ can appear in $G^\pm$ in (\ref{eq:Gamma}) only
in the terms $(A_t)_k (F_t)_{h,k} (A_t)_k \rightarrow (A_t)_k
(F_t)_{h,k} (A_t)_k -t d\eta [(F_t)_{h,k},(A_t)_k]$. Since we must
take the trace of this with a Goldstone boson $\xi$ in $G/H$ in order
to get a non-vanishing contribution via the anomaly, and since the
generators in $\mathfrak{g}$ are orthogonal, the sole such
contribution to the action is given by 
\begin{gather}
\Gamma \supset 4c_+ \int t d \eta \mathrm{tr} \xi [(F_t)_h,(A_t)_k].
\end{gather}
In addition, we get contributions where we take terms in $G^\pm$ not
involving $\eta$, of the form 
\begin{gather}
\Gamma \supset \sum_\pm c_\pm \int  \eta \mathrm{tr} G^\pm[A_t].
\end{gather}
These simplify dramatically. Indeed, orthogonality of generators,
together with $\mathrm{tr}
(A_k^2 F_k +A_kF_kA_k + F_k A_k^2)= \mathrm{tr}
A_k^2 F_k = 0$, implies that $G^-
=0$. Moreover, since $\mathrm{tr} (A_t)_k^4 = 0$, we see that there
can be no WZW term arising from our anomalous effective action in the
ungauged limit. 

All in all, we find that the anomalous action can be simplified to 
\begin{gather}
\Gamma = c_+ \int \left( \eta \mathrm{tr} [3(F_t)_h^2 + (F_t)_k^2 -4
  (A_t)_k^2(F_t)_h] + 4t d \eta \mathrm{tr} \xi [(F_t)_h,(A_t)_k]\right).
\end{gather}

We now consider the contributions of each of the triangle anomalies in turn. For $SU(3)^2
U(1)$ and the anomalies involving $U(1)$s, the effective
action just reduces to 
\begin{gather}
\int c_3 \eta \mathrm{tr} GG + c_1 \eta BB
\end{gather}
to all orders. 

Things are somewhat more complicated for the $SO(5)^2 U(1)$ anomaly.
Let us content ourselves with computing the action at the
next-to-leading order. Evidently, we have that
\begin{align}
F_k &= t[\xi, F] + \dots \\
F_h &= F +\frac{t^2}{2} [\xi,[\xi,F]] + \dots \\
A_k &= -t (d\xi - [\xi,A]) +\dots  \equiv -t (D\xi) +\dots.
\end{align}
From which it is clear that the first corrections arise not at
dimension 6, but at dimension 7. Explicitly, we find\footnote{The term
  multiplied by $\frac{1}{2}$ simplifies to $[[F^2,\xi],\xi]$, but
  we prefer to write it in a form that leaves the Lie algebra structure manifest.}
\begin{gather}
\int \frac{c_5}{3} \eta \mathrm{tr}
(3F^2+\frac{1}{2}(F[\xi,[\xi,F]]+[\xi,[\xi,F]]F + 2[\xi,F]^2)-\frac{4}{3} (D\xi)^2F) +\frac{4}{3} d\eta \mathrm{tr} \xi [
F,D\xi] +\dots
\end{gather}
The last term may be integrated by parts, to get 
\begin{gather}
\int \frac{4}{3} d\eta \mathrm{tr} \xi [
F,D\xi] = \int \frac{4}{3} \eta \mathrm{tr} (2(D\xi)^2 F
-\xi[F,[F,\xi]]).
\end{gather}
Finally, we obtain
\begin{gather}
\int \frac{c_5}{3} \eta \mathrm{tr}
(3F^2+\frac{1}{2}(F[\xi,[\xi,F]]+[\xi,[\xi,F]]F + 2[\xi,F]^2)+\frac{4}{3} (D\xi)^2F) -\frac{4}{3}\xi[F,[F,\xi]] +\dots
\end{gather}
To convert this into an explicit formula in terms of
$SU(2) \times U(1)$ invariant operators in the basis of \cite{Gripaios:2016xuo}, we use the basis for $\mathfrak{so}(5) \simeq
\mathfrak{sp}(2)$ \cite{Gripaios:2009pe}, wherein \footnote{We have removed erroneous
  factors of $\pm i$ that appear in \cite{Gripaios:2009pe}.}  
\begin{gather} \label{eq:basis}
F = \frac{1}{2}\begin{pmatrix} W^i \sigma^i & 0 \\ 0 & B
  \sigma^3 \end{pmatrix}, \xi = \begin{pmatrix} 0 & (H^c H) \\ (H^c
  H)^\dagger &0\end{pmatrix}.
\end{gather}
The only non-vanishing term at next-to-leading order is the last one,
for which
\begin{gather}
\mathrm{tr} \xi[F,[F,\xi]] = H^\dagger H (W^i W^i + B^2) + 2 H^\dagger
\sigma^i H W^i B,
\end{gather}
where $W^i$ and $B$ are the field strength 2-forms.

Putting everything together, we obtain the expression in eq. \ref{eq:effactho}.

\bibliographystyle{JHEP}
\bibliography{references}

\end{document}